%% file: main.tex
\def\Version{5.2} 




\message{<< Assuming 8.5" x 11" paper >>}    

\magnification=\magstep1	          
\raggedbottom
\parskip=9pt

\def\singlespace{\baselineskip=12pt}      
\def\sesquispace{\baselineskip=16pt}      



\input mathmacros
\input mathmacros.greekbold
\input msmacros


\def\sqr#1#2{\vcenter{
  \hrule height.#2pt 
  \hbox{\vrule width.#2pt height#1pt 
        \kern#1pt 
        \vrule width.#2pt}
  \hrule height.#2pt}}

\def\dal{\mathop{\,\sqr{7}{5}\,}}
\def\block{\dal} 

\def\boxK{\dal\limits_{K} \phantom{}}

\def\region{\Buchstabe{X}}
\def\interval#1#2{\langle #1, \, #2 \rangle}
\def\Exp{\Euclid}
\def\Bbar{\bar{B}}
\def\f{\phi}

\def\IR{R}
\def\UV{l}
\def\meso{\lambda_0}
\def\Mink{\Minkowski}

\def\O{{\cal O}}


\input epsf
\epsfverbosetrue

\def\FigureNumberCaption#1#2#3{
 \singlespace
 \vbox{
   \centerline{\vbox{\epsfbox{#1}}}		
   \leftskip=1.5truecm\rightskip=1.5truecm	
     \vskip 15pt
     \noindent{\it Figure #2}. #3		
     \vskip .25in
   \leftskip=0truecm\rightskip=0truecm}		
 \vfill 
 \sesquispace}
 %


\phantom{}

\PrintVersionNumber

\sesquispace
\centerline{
  {\titlefont Does Locality Fail at Intermediate Length-Scales?}%
  \footnote{$^{^{\displaystyle\star}}$}%
{To appear in
 Daniele Oriti (ed.),
 {\it Towards Quantum Gravity} 
 (Cambridge University Press, 2007)}} 

\bigskip


\singlespace			        

\author{Rafael D. Sorkin}
\address
 {Perimeter Institute, 31 Caroline Street North, Waterloo, ON N2L 2Y5, Canada}
\furtheraddress
 {Department of Physics, Syracuse University, Syracuse, NY 13244-1130, U.S.A.}
\email{sorkin@physics.syr.edu}

\AbstractBegins                              
If quantum gravity implies a fundamental spatiotemporal discreteness,
and if its ``laws of motion'' are compatible with the Lorentz
transformations, then physics cannot remain local.  One might expect 
this nonlocality to be confined to the fundamental discreteness
scale, but I will present evidence that it survives at much lower
energies, yielding for example a nonlocal equation of motion for a
scalar field propagating on an underlying causal set.
\AbstractEnds


\sesquispace


\noindent
Assuming that ``quantum spacetime'' is fundamentally discrete, how
might this discreteness show itself?  
Some of its potential
effects are more evident, others less so.  The atomic and molecular
structure of ordinary matter influences the propagation of both waves
and particles in a material medium.  Classically, particles can be
deflected by collisions and also retarded in their motion, giving rise in
particular to viscosity and Brownian motion.  In the case of
spatio-temporal discreteness, 
viscosity is excluded by
Lorentz symmetry, but fluctuating deviations from rectilinear
motion are still possible.  Such ``swerves'' have been described in
[1] and [2].  They depend (for a massive particle)
on a single phenomenological parameter, essentially a diffusion constant
in velocity space.  
As far as I know, the corresponding analysis for a
quantal particle with mass has not been carried out yet, 
but for massless quanta such as photons 
the diffusion equation of [1] can be adapted to 
say something, and it then describes 
fluctuations of both energy and polarization 
(but not of direction),
as well as a secular ``reddening'' (or its opposite).  
A more complete quantal
story, however, would require that particles be treated as wave
packets, 
raising the general question of how spatiotemporal
discreteness affects the propagation of {\it waves}.  
Here, the analogy with
a material medium
suggests 
effects such as scattering and extinction,
as well as possible nonlinear effects. 
Further generalization to a ``second-quantized field'' might have more
dramatic, if less obvious, consequences.  
In connection with cosmology,
for example,  
people have wondered how discreteness
would affect the hypothetical inflaton field.

So far, I have been assuming that, although the deep structure of
spacetime is discrete, it continues to respect the Lorentz
transformations.  
That this is logically possible is demonstrated [3] by the
example of causal set (causet) theory [4].
With
approaches such as loop quantum gravity, 
on the other hand,
the status of local Lorentz
invariance seems to be controversial.  Some people have hypothesized
that it would be broken or at least ``deformed'' in such a way that the
dispersion relations for light would cease to be those of a massless
field.  Were this the case, empty space could also resist the passage of
particles (a viscosity of the vacuum), since there would now be a state
of absolute rest.   Moreover, reference [5] has argued
convincingly that  it would be difficult to avoid $O(1)$
renormalization effects that would lead to different quantum fields
possessing different effective light cones.  Along these lines, one might
end up with altogether more phenomenology than one had bargained for.

As already mentioned, the causal set hypothesis avoids such
difficulties, but in order to do so, it has to posit a kinematic
randomness, in the sense that a spacetime\footnote{$^\star$}
{In this article, ``spacetime'' will always mean Lorentzian
 manifold, in particular a continuum.}
$M$ 
may
properly correspond 
only to causets $C$ that could have been produced by
a 
{\it Poisson process} in $M$.  With respect to
an approximating spacetime $M$, 
the causet thus functions as
a kind of ``random lattice''.  
Moreover, 
the infinite volume of
the Lorentz group implies that such a ``lattice'' cannot be home to a
local dynamics.  Rather the ``couplings'' or ``interactions'' that
describe physical processes occurring in the causet are 
--- of necessity ---
radically nonlocal.

To appreciate why this must be, let us refer to the process that will be
the subject of much of the rest of this paper: propagation of a scalar
field $\f$ 
 on a background causet $C$
that is  well approximated by a Minkowski
spacetime $M=\Minkowski^d$.  To describe 
such a
dynamics, 
one needs
to reproduce within $C$ something like the d'Alembertian operator
$\block$, the Lorentzian counterpart of the Laplacian operator 
$\nabla^2$
of
Euclidean space $\Euclid^3$.  Locality in the discrete context, if it
meant anything at all, would imply that 
the action of $\dal$
would be built up
in terms of ``nearest neighbor couplings''
(as in fact $\nabla^2$ can be built up, 
on either a crystalline or random lattice in $\Euclid^3$).
But Lorentz invariance 
contradicts this sort of locality
because it implies that, no
matter how one chooses to define nearest neighbor, any given causet
element $e\in C$ will possess an immense number of them 
extending throughout 
the region of $C$ corresponding to the light cone of
$e$ in $M$.  
In terms of a Poisson process in $M$ we can 
express this more precisely by saying 
that the {\it probability} of any given element $e$ possessing a
limited
 number of
nearest neighbors is vanishingly small.  Thus, the other
elements to which $e$ must be ``coupled'' by our box operator 
will be large
in number (in the limit infinite), and in any given frame of reference,
the vast majority of them 
will be
remote
from $e$.
The resulting ``action at a distance'' epitomizes the maxim
 that discreteness plus Lorentz invariance entails
nonlocality.

If this reasoning is correct, it implies that physics at the Planck
scale must be radically nonlocal.  
(By Planck scale I just mean the fundamental length- or volume-scale
associated with the causet or other discrete substratum.)  
Were it to be confined to the Planck scale,
however,
this nonlocality would be of limited phenomenological interest
despite its deep significance for the underlying theory.  
But a little thought
indicates that things might not be so simple.  
On the contrary,
it is far from obvious that 
the kind of nonlocality in question 
can be confined to any scale, 
because
for any given configuration of the field $\phi$, the  ``local
couplings'' will be vastly outnumbered by the ``nonlocal'' ones.  How
then could the latter conspire to cancel out so that the former could
produce a good approximation to $\block\phi$,
even for a slowly varying $\phi$?  

When posed like this, the
question looks almost hopeless, but I will try to convince you that
there is in fact an answer.  
What the answer seems to say,
though, 
is that
one can reinstate locality only conditionally and to a limited extent. 
At any finite scale $\lambda$,
some nonlocality will naturally
persist, 
but the scale $\meso$
at which it begins to disappear 
seems to 
reflect
not
only the ultraviolet scale $\UV$ 
but also an infrared scale $\IR$,
which we may identify with the age of the cosmos,
and which
(in a kind of quantum-gravitational echo of Olber's paradox) 
seems to be needed in order that locality 
be recovered at all.  
On the other hand, (the) spacetime (continuum) as such can make sense
almost down to $\lambda=\UV$.
We may thus anticipate
that, 
as we coarse-grain up from $\UV$ to larger and larger sizes $\lambda$,
we will reach 
a stratum of reality in which
 discontinuity has faded out and
spacetime has emerged,
but physics continues to be nonlocal.  
 One would
expect 
 the best description
of this stratum to be some type of nonlocal field theory;  
and this would be a
new sort of manifestation of discreteness: 
not as a source of
fluctuations, but as a source of nonlocal phenomena.

Under still further coarse-graining, this nonlocality should disappear as
well, and one might think that one would land for good in the realm of
ordinary quantum field theory (and its further coarse-grainings).
However, there is reason to believe that locality would fail once again
when cosmic dimensions were reached;
in fact, 
the non-zero cosmological constant predicted on the basis of causet theory 
is very much a 
nonlocal reflection, 
on the largest scales,
of the underlying discreteness.
It 
is a strictly quantal effect,
however, 
and would be a very
different sort of residue of microscopic discreteness
than what I'll be discussing here.

These introductory remarks express in a general way most of what I want to
convey in this paper, but before getting to the technical
underpinnings, let me just (for shortage of space) list some other
reasons why people have wanted to give up locality as a fundamental
principle of spacetime physics: 
to cure the divergences of quantum field theory (e.g. [6]);
to obtain particle-like excitations of a spin-network or related graph
[7];
to give a realistic and deterministic account of quantum mechanics 
(the Bohmian interpretation is both nonlocal and acausal, for example);
to let information escape from inside a black hole (e.g. [8]);
to describe the effects of hidden dimensions in ``brane world'' scenarios;
to reduce quantum gravity to a flat-space quantum field theory via the
so called AdS-CFT correspondence;
to make room for non-commuting spacetime coordinates.
(This  ``non-commutative geometry'' reason is perhaps the most
suggestive in the present context, because it entails a hierarchy
of scales analogous to the scales $\UV$, $\meso$ and $\IR$.  On the ``fuzzy
sphere'' in particular, the non-commutativity scale $\meso$ is the
geometric mean between the 
effective ultraviolet cutoff $\UV$ and the sphere's radius $\IR$.)


\section{Three D'Alembertians for two-dimensional causets}
The scalar field on a causet offers a simple model for the questions we
are considering.  Kinematically, we may realize such a field simply as a
mapping $\phi$ of the causet into the real or complex numbers, 
while in the continuum
its equations of motion take 
--- at the classical level ---
the simple form $\block\phi=0$,
assuming (as we will) that the mass vanishes.  In order to make sense of
this equation in the causet, 
we ``merely'' need to give a
meaning to the D'Alembertian operator $\block$.  This is not an easy
task, but it seems less difficult than  giving meaning to,
for example, the gradient of $\phi$ (which for its accomplishment would
demand that we define  a concept of vectorfield on a causet).  Of
course, one wants ultimately to treat the quantum case, but one would
expect a definition of $\block$ to play a basic role there as well, so
in seeking such a definition we are preparing equally for the classical
and quantal cases.

If we assume that $\block$ should act linearly on $\phi$ 
(not as obvious as one might think!), 
then our task reduces to the finding of a suitable matrix $B_{xy}$
to play the role of $\dal$,
where the indices
$x$, $y$
range over the elements of the causet $C$.
We will also require that $B$ be ``retarded'' or ``causal'' in the sense
that $B_{xy}=0$ whenever $x$ is spacelike to, or causally precedes $y$.
In the first place, this is helpful classically, since it allows one to
propagate a solution $\phi$ forward iteratively, element by element
(assuming that the diagonal elements $B_{xx}$  do not vanish).
It might similarly be advantageous quantally, if the path integration
is to be conducted in the Schwinger-Kel'dysh manner.

\subsection {First approach through the Green function}
I argued above that no matrix $B$ that (approximately) respects the Lorentz
transformations can reproduce a local expression like the
D'Alembertian unless the majority of terms cancel miraculously 
in the sum, 
\ $\sum\limits_{y} B_{xy}\phi_y =: (B\phi)_x$~,
that corresponds to $\dal\phi(x)$.

Simulations by Alan Daughton [9],
continued by Rob Salgado [10],
provided the  first
evidence that the required cancellations can actually be
arranged for 
without
appealing to anything other than the intrinsic order-structure of the
causet.  In this approach one notices that, although in the natural
order of things one begins with the D'Alembertian and ``inverts'' it to
obtain its Green function $G$, the result in $1+1$-dimensions is so
simple that the procedure can be reversed.  In fact, the {\it retarded}
Green function $G(x,y)=G(x-y)$ in $\Minkowski^2$ is 
(with the sign convention $\block=-\ptl^2/\ptl{t}^2+\ptl^2/\ptl{x}^2$) 
just the step function with magnitude $-1/2$ and support the future
of the origin (the future light cone together with its interior).
Moreover, thanks to the conformal invariance of $\block$
in $\Mink^2$,
the same expression remains valid in the presence of spacetime
curvature. 

Not only is this continuum expression very simple, 
but it has an obvious counterpart in the causal set,
since it depends on nothing more than the causal relation between the two
spacetime points $x$ and $y$.
Letting the symbol $<$ denote (strict) causal precedence in the usual
way, we can represent the causet $C$ as a matrix whose elements
$C_{xy}$ take the value $1$ when $x<y$ and $0$ otherwise.  
The two-dimensional analog  $G$ of the retarded Green function  is then
just $-1/2$ times (the transpose of) this matrix. 

From these ingredients, one can concoct some obvious candidates for the
matrix $B$.  
 The one that so far has worked best is obtained
by symmetrizing $G_{xy}$ and then inverting it.
More precisely, what has been done is the following:
begin with a specific region $R\subset\Minkowski^2$ 
(usually chosen to be an order-interval, the diamond-shaped region lying
causally between a timelike pair of points);
randomly sprinkle $N$ points $x_i,\ i=1\dots N$ into $R$;
let $C$ be the causet with these points as substratum and the
order-relation $<$ 
induced from $\Minkowski^2$;
for any 
``test''
scalar field $\phi$ on $R$, 
let $\phi_i=\phi(x_i)$ be the induced ``field'' on $C$;
build the $N\times N$ matrix $G$ and then symmetrize and invert to get
$B$, as described above; 
evaluate $B(\phi,\psi)=\sum_{ij}B_{ij}\phi_i\psi_j$
for $\phi$ and $\psi$ drawn from a suite of test functions
on $R$; compare with the continuum values,
$\int d^2x \, \phi(x)\block\psi(y) \, d^2y$.   

For test functions that vanish to first order on the boundary
$\partial{R}$ of $R$, and that vary slowly on the scale set by the
sprinkling density, the results so far exhibit full agreement between
the discrete and continuum values [9][10].  
Better agreement than this, one could not have hoped for in either
respect:
Concerning boundary terms, the heuristic reasoning that
leads one to expect that inverting a Green function will reproduce
a discretized version of
$\block$ leaves open 
its behavior on $\partial R$. 
Indeed, one doesn't really know what
continuum expression to compare with:
If our fields don't vanish on $\partial R$, 
 should we expect to obtain an approximation to
 $\int dx dy \phi(x)\block\psi(y)$ or
 $\int dx dy (\grad\phi(x),\grad\psi(y))$ or \dots?
Concerning rapidly varying functions, it goes without saying that, just
as a crystal cannot support a sound wave shorter than the interatomic
spacing, a causet cannot support a wavelength shorter than $\UV$.  But
unlike with crystals, this statement requires some qualification because
the notion of wavelength is frame-dependent.  What is red light for
one inertial observer is blue light for another.  Given that the
causet can support the red wave, it must be able to support the blue one
as well, 
assuming  Lorentz invariance in a suitable sense. 
Conversely, such 
paired fields can be used to test the Lorentz invariance of 
$B$.  To the limited extent that this important test has been done, 
the results have also been favorable. 

On balance, then, the work done on the Green function approach gives
cause for optimism that ``miracles do happen''.
However, the simulations have been limited to the flat case, and, more
importantly, they do not suffice (as of yet) to establish that the
discrete D'Alembertian $B$ is truly frame independent.  
The point is
that although $G$ itself clearly is Lorentz invariant in this sense, its
inverse (or rather the inverse of the symmetrized $G$) 
will in general depend on the region $R$ in which one works. 
Because this region is not itself invariant under boosts, it defines a
global frame that could find its way into the resulting matrix $B$.
Short of a better analytic understanding, one is unable
to rule out this subtle sort of frame dependence, 
although the
aforementioned limited tests provide evidence against it.

Moreover, the Green function prescription itself is of limited
application.  In addition to two dimensions, the only other case where a
similar prescription is known is that of four dimensions {\it without}
curvature, where one can take for $G$ the ``link matrix'' instead of the
``causal matrix''.

Interestingly enough, the potential for Lorentz-breaking by
the region $R$ does not arise if one works exclusively with retarded
functions, that is, if one forms $B$ from  the original retarded
matrix $G$, rather than its
symmetrization.\footnote{$^\dagger$}
{One needs to specify a nonzero diagonal for $F$.}
Unfortunately, 
computer tests with the retarded Green function 
have so far been
discouraging
on the whole
(with some very recent exceptions).  
Since, for 
quite
different reasons, it would be
desirable to find a retarded representation of $\block$, 
this suggests
that we try something different.

\subsection {Retarded couplings along causal links}
Before taking leave of the Green function scheme just described, we can
turn to it for one more bit of insight.  If one examines the individual
matrix elements $B_{xy}$
for a typical sprinkling, one notices
first of all that they seem to be equally distributed among positive and
negative values, and second of all that the larger magnitudes among them
are concentrated ``along the light cone''; that is, $B_{xy}$ tends to be
small unless the  proper distance between $x$ and $y$ is near zero.
The latter observation may remind us of a collection of ``nearest
neighbor couplings'', here taken in the only possible Lorentz invariant
sense: that of small proper distance.  The former observation suggests 
that a recourse to oscillating signs might be the way to effect the
``miraculous cancellations'' we are seeking.

The suggestion of oscillating signs is in itself rather vague, but two
further observations will lead to a more quantitative idea,
as illustrated in figure 1.
Let $a$ be
some point in $\Minkowski^2$, let $b$ and $c$ be points on 
the right and
left halves of its past lightcone 
(a ``cone'' in $\Mink^2$ being just a pair of null rays), 
and
let $d$ be the fourth point needed to complete the rectangle.  If (with
respect to a given frame) all four points are chosen to make a small
square, and if $\phi$ is slowly varying (in the same frame), then the
combination $\phi(a)+\phi(d)-\phi(b)-\phi(c)$ converges, after suitable
normalization, 
to \ $-\block\phi(a)$ 
as the square 
shrinks to zero size.
(By
Lorentz invariance, the same would have happened even if we had started
with a rectangle rather than a square.)  
On the other hand, four other
points obtained from the originals by a large boost will form a long
skinny rectangle, in which  the points $a$ and $b$ (say) 
are very close
together, as are $c$ and $d$.  Thanks to the profound identity,
$\phi(a)+\phi(d)-\phi(b)-\phi(c)=\phi(a)-\phi(b)+\phi(d)-\phi(c)$, we
will obtain only a tiny contribution from this rectangle --- exactly the
sort of cancellation we were seeking!  
By including all the boosts of
the original square,
we might thus hope to do justice to the Lorentz
group without bringing in the unwanted contributions we have been
worrying about.  


Comparison with
 the D'Alembertian in  
one dimension leads to a
similar idea, 
which in addition works a bit better in the causet, where
elements corresponding to the type of ``null rectangles'' just discussed
don't really exist.  In $\Minkowski^1$, which is just the real line, 
$\block\phi$ reduces (up to sign) to $\ptl^2\phi/\ptl{t}^2$, for which
a well known discretization is $\phi(a)-2\phi(b)+\phi(c)$, $a$, $b$
and $c$ being three evenly spaced points.  
Such a configuration {\it does} find correspondents in the causet,
for example 3-chains $x<y<z$ such that no element other than $y$ lies
causally between $x$ and $z$
Once again, any single one of these chains (partly) determines a frame,
but the collection of all of them does not.  
Although these examples should not be taken too seriously 
(compare the sign in equation (1) below),
they bring us very close
to the following scheme.\footnote{$^\flat$}
{A very similar idea was suggested once by Steve Carlip}

Imagine a causet $C$ consisting of points sprinkled into a region of
$\Mink^2$, and fix an element $x\in C$ at which we would like to know
the value of $\block\phi$.  We can divide the ancestors of $x$ (those
elements that causally precede it) into ``layers'' according to their
``distance from $x$'', as measured by the number of intervening
elements.  Thus layer 1 comprises those $y$ which are
{\it linked} 
to $x$ in the sense that $y<x$ with no intervening elements,
 layer 2 comprises those $y<x$ with only a single element
$z$ such that  $y<z<x$, etc.
(Figure 2 illustrates the definition of the layers.)   
Our prescription for $\block\phi(x)$ is
then to take some combination, with alternating signs, of the first few
layers, the specific coefficients to be chosen so that the correct
answers are obtained from suitably simple test functions.  
Perhaps the simplest combination of
this sort is
$$
   B\phi(x)
   =
   {4\over l^2} 
   \left(
     -\half\phi(x) + \left( \sum_1 - 2 \sum_2 + \sum_3 \right) \phi(y)
   \right)
  \eqno(1)
$$
where the three sums $\sum$ extend over the first three layers 
as just defined, and $l$ is the fundamental length-scale associated with
the sprinkling, normalized so that each sprinkled point occupies, on
average, an area of $l^2$.
The prescription (1) yields a candidate for the
``discrete D'Alembertian''
$B$
which 
is {\it retarded}, 
unlike 
our earlier
candidate based on the symmetrized Green function.
In order to express this new $B$ explicitly as a matrix,
let $n(x,y)$ denote the cardinality of
the order-interval $\interval{y}{x}=\braces{z\in C| y < z < x}$, 
or in
other words the number of elements of $C$ causally between $y$ and $x$.
Then, 
assuming that $x\ge y$,
we have from (1),
\def\AA{{l^2 \over 4}B_{xy}}
\def\BB{-1/2 \quad {\rm\ for\ } x=y}
\def\CC{1, -2, 1, 
     \ {\rm  according \ as\ } n(x,y) {\rm \ is\ } 0, 1, 2, {\rm\ respectively} ,
	\quad {\rm\ for\ } x\not=y}
\def\DD{\ \ 0 \qquad {\rm\ otherwise}}
$$
   \AA = \left\{  \hbox{$\eqalign{&\BB \cr &\CC \cr &\DD \cr}$} \right.
   \eqno(2)
$$

Now let $\phi$ be a fixed test function of compact support on $\Mink^2$,
and let $x$ 
(which we will always take to be included in $C$)
be a fixed point of $\Mink^2$.
If we apply $B$ to $\phi$ we will of course obtain a random
answer depending on the random sprinkling of $\Mink^2$.
However, 
one can prove that the {\it mean} 
of this random variable,
$\Exp B\phi(x)$,
converges to 
$\block\phi(x)$ in the continuum limit $l\to0$:
$$
 \Exp \  \sum_y B_{xy}\phi_y 
 \quad
 \to\limits_{l\to0}
 \quad
 \dal\phi(x)  
 \ ,
 \eqno(3)
$$
where $\Exp$ denotes expectation with respect to the Poisson process
that generates the sprinkled causet $C$.  
[The proof rests on the following facts.  Let us limit the sprinkling to
an ``interval'' (or ``causal diamond'') $\region$ with $x$ as its top
vertex.  For test functions that are polynomials of low degree, one can
evaluate the mean in terms of simple integrals over $\region$ ---
for example the integral 
$\int {dudv/l^2} \exp\braces{-uv/l^2} \ \phi(u,v)$ ---
and the
results agree with $\block\phi(x)$, up to corrections that vanish like
powers of $l$ or faster.]

In a sense, then, we have successfully reproduced the D'Alembertian in
terms of a causet expression that is {\it fully intrinsic} and therefore
automatically {\it frame-independent}.   Moreover, the matrix $B$,
although it introduces nonlocal couplings, does so only on Planckian
scales, which is to say, on scales no greater than demanded by the
discreteness itself.\footnote{$^\star$}
{It is not difficult to convince oneself that 
 the limit in  (3) sets in 
 when $\UV$ shrinks below the characteristic length associated with the
 function $\phi$; 
 or vice versa, if we think of $\UV$ as fixed, 
 $B\phi$ will be a good approximation to $\dal\phi$ when the
 characteristic length-scale $\lambda$ over which $\phi$ varies exceeds
 $\UV$: $\lambda \gg \UV$.
 But this means in turn that 
 $(B\phi)(x)$ can be sampling $\phi$ 
 {\it in effect} 
 only in a neighborhood of $x$ 
 of characteristic size $\UV$. 
 Although $B$ is thoroughly nonlocal at a fundamental level,  
 the scale of its effective nonlocality in application to slowly varying
 test functions is thus no greater than $\UV$.}

But is our ``discrete D'Alembertian'' $B$ really a satisfactory 
tool for building a field theory on a causet?  The potential problem
that suggests the opposite
conclusion
 concerns the fluctuations in (1),
which grow with $N$ rather than dying away.  (This growth is indicated
by theoretical estimates and confirmed by numerical simulations.)
Whether this problem is fatal or not is hard to say.  For example, in
propagating a classical solution $\phi$ forward in time through the
causet, it might be that the fluctuations in $\phi$ induced by those in
(1) would remain small when averaged over 
many
Planck lengths, so
that the coarse-grained field would not see them.  But if this is true,
it remains to be demonstrated.  And in any case, the fluctuations would
be bound to affect even the coarse-grained field when they became big
enough.  For the remainder of this paper, I will assume that large
fluctuations are not acceptable, and that one
consequently
needs a 
different
$B$ that will yield the desired answer not only on average,
but (with high probability) in each given case.  For that purpose, 
we will have to make more complicated
the
remarkably simple ansatz (2) that we arrived at above.

\subsection {Damping the fluctuations}
To that end, 
let us return to 
equation
(3)
and notice that 
$\Exp(B\phi)=(\Exp{B})\phi$, 
where
what I have
just
 called
$\Exp{B}$ 
is effectively a continuum integral-kernel $\bar{B}$
in $\Mink^2$.
That is to say, 
when we average over all sprinklings
to get 
$\Exp B\phi(x)$, 
the sums in (1)
turn into 
integrals and there results an expression of the form 
$\int\bar{B}(x-y)\phi(y)d^2y$,
where $\bar{B}$ is a retarded, continuous function
that can be computed explicitly.
Incorporating into $\Bbar$ 
the $\delta$-function  answering to $\phi(x)$ in
(1),
we get for our kernel (when $x>y$),
$$
  \Bbar (x-y) = 
  {4\over l^4} \, p(\xi)e^{-\xi} \; - \; {2\over l^2} \delta^{(2)}(x-y) \ ,
 \eqno(4)
$$
where 
$p(\xi)=1-2\xi+\half\xi^2$, 
$\xi=v/l^2$ 
and
$v=\half||x-y||^2$
is the volume (i.e. area) of the order-interval in $\Mink^2$ delimited
by $x$ and $y$.
The convergence result (3) then states that,
for $\phi$ of compact support,
$$
 \int \Bbar(x-y)\phi(y) d^2y 
 \quad
 \to\limits_{l\to0}
 \quad
 \dal\phi(x)  
 \eqno(5)
$$
Notice that, as had to happen,
$\Bbar$ is Lorentz-invariant, 
since it depends only on the invariant interval 
$||x-y||^2=|(x-y)\cdot(x-y)|$.\footnote{$^\dagger$}
{The existence of a Lorentz-invariant kernel $\bar{B}(x)$ that yields
 (approximately) $\dal\phi$ might seem paradoxical, 
 because one could take 
 the
 function $\phi$ itself to be Lorentz invariant (about the origin $x=0$, say),
 and
 for such a $\phi$ the integrand in (5)
 would also be invariant,
 whence the integral 
 would apparently have to diverge.
 This divergence 
 is avoided for compactly supported $\phi$, 
 of course, because the potential divergence is cut off where the
 integrand goes to zero. 
 But 
 what is truly remarkable in the face of the 
 counter-argument just given,
 is that 
 the answer is insensitive to the size of the supporting region.
 With any reasonable cutoff
 and reasonably well behaved test functions,
 the integral still manages to converge to the correct answer 
 as the cutoff is taken to infinity.
 Nevertheless, this risk of divergence hints at the need
 we will soon encounter for some sort of infrared cutoff-scale.}

Observe, now, that the fundamental discreteness-length has all but
disappeared from our story.  It remains only in the form of a parameter
entering into the definition (4) of the integral kernel
$\Bbar$.  As things stand, this parameter reflects the scale of
microscopic physics from which $\Bbar$ has emerged (much as the diffusion
constants of hydrodynamics reflect
atomic dimensions).
But nothing in the definition of $\Bbar$ per se forces us to this
identification.  
If in (4) we replace $l$ by
a freely variable length,
and if we then follow the Jacobian dictum,
``Man muss immer umkehren''\footnote{$^\flat$}
{``One must always reverse direction.''}
we can arrive at a modification of the discrete D'Alembertian $B$ for
which the unwanted fluctuations are damped out by the law of large
numbers. 

Carrying out the first step, let us replace $1/\UV^2$ in (4)
by
a new
parameter $K$.  We obtain 
a new  continuum approximation 
to $\dal$,
$$
  \Bbar_K (x-y) = 
  {4 K^2} \, p(\xi)e^{-\xi} \; - \; {2K} \delta^{(2)}(x-y) \ ,
 \eqno(6)
$$
whose associated nonlocality-scale is not $\UV$ but the length
$K^{-1/2}$, which we 
can 
take to be much larger than $\UV$.
%
Retracing the steps that led from the discrete matrix
(2) to the continuous kernel (4) 
then
brings us to the following
causet
expression that yields 
(6) when its
sprinkling-average is taken:
$$
   B_K\phi(x)
   =
   {4\eps\over l^2}
   \left(
    - \half \phi(x) 
    + \eps \sum_{y < x} f(n(x,y),\eps) \  \phi(y)
   \right) \ ,
  \eqno(7)
$$
where 
$\eps=l^2K$,
and
$$
   f(n,\eps) = (1-\eps)^n 
   \left(
     1 - {2\eps n\over 1-\eps} + {\eps^2 n (n-1)\over 2 (1-\eps)^2}
   \right) 
   \ .
   \eqno(7a)
$$
For $K=1/l^2$ we recover (1).
In the limit where $\eps\to0$ and $n\to\infty$, 
$f(n,\eps)$
reduces to
the now familiar form 
$
  p(\xi) e^{-\xi}
$
with $\xi=n\eps$. 
That is, we obtain in this limit  
the Montecarlo approximation 
to the integral $\Bbar_K\f$
induced by the sprinkled points.
(Conversely,
$p(\xi)e^{-\xi}$
can serve as a lazybones' alternative 
to (7a)).

Computer
simulations show that $B_K\phi(x)$ furnishes a good approximation to 
$\dal\phi(x)$ for simple test functions, but this time 
one finds
 that
the fluctuations {\it also} go to zero with $l$, assuming the
physical nonlocality scale $K$ remains fixed as $l$ varies.  
For example, with $N=2^9$ points sprinkled into the interval in
$\Mink^2$ delimited by $(t,x)=(\pm1,0)$,
and with the test functions 
$\f=1, t, x, t^2, x^2, tx$,
the fluctuations 
in $B_K\f(t=1,x=0)$
for $\eps=1/64$
range 
from a standard deviation 
of $0.53$ (for $\f=x^2$)
to $1.32$ (for $\f=1$);
%
%
and they die out roughly like $N^{-1/2}$ 
(as one might have expected) 
when $K$ is held fixed as $N$ increases.
The means are accurate by construction,
in the sense that they
exactly\footnote{$^\star$}
{Strictly speaking, this assumes that the number of sprinkled points is Poisson
 distributed, rather than fixed.}
reproduce the continuum expression $\Bbar_K\f$
(which in turn reproduces $\dal\f$ to an accuracy of around 1\%
for $K\gto 200$.)
(It should also be possible to estimate the fluctuations analytically, 
but I have not tried to do so.)

In any case, 
we can conclude that ``discretized D'Alembertians'' 
suitable for causal sets do exist, a fairly simple one-parameter family
of them being given by (7). 
The parameter $\eps$ in that expression determines the 
scale of the nonlocality 
via $\eps=Kl^2$, and it must be $\ll1$ if we want the fluctuations in
$B\f$ to be small.
In other words, we need a significant separation between the two 
length-scales $l$ and 
$\meso=K^{-1/2}=l/\sqrt{\eps}$.

\section {Higher dimensions}
So far, we have been concerned primarily with two-dimensional causets
(ones that are well approximated by two-dimensional spacetimes).
Moreover, the quoted result,
(3) cum (6),
has been proved
only under the 
additional assumption of flatness, 
although it seems likely that it could be extended to the
curved case.  More important, however, is finding D'Alembertian
operators/matrices for four and other dimensions.  
It turns out that one can do this systematically in a way that
generalizes what we did in two dimensions.

Let me illustrate the underlying ideas in the case of 
four dimensional Minkowski space $\Mink^4$.
In $\Mink^2$ we began with the D'Alembertian matrix $B_{xy}$, averaged
over sprinklings to get 
$\Bbar(x-y)$, and ``discretized'' a rescaled $\Bbar$ to get the matrix 
$(B_K)_{xy}$. 
It turns out that this same procedure works in 4-dimensions if we begin
with the coefficient pattern $1$~$-3$~$3$~$-1$ instead of $1$~$-2$~$1$.

To see why it all works, however, it is better to start with the
integral kernel and not the matrix
(now that we know how to pass between them).  
In $\Mink^2$ we found $\Bbar$ in the form of a delta-function plus a term
in $p(\xi)\exp(-\xi)$, where
$\xi=Kv(x,y)$, 
and $v(x,y)$ was the volume of the order-interval $\interval{y}{x}$,
or equivalently --- in $\Mink^2$ --- Synge's ``world
function''.
In other dimensions this equivalence breaks down and we can imagine using
either the world function or the volume (one being a simple power of the
other, up to a multiplicative constant).
Whichever one chooses, the real task is to find the polynomial $p(\xi)$
(together with the coefficient of the companion delta-function term).

To that end, notice that the combination $p(\xi)\exp(-\xi)$ can always
be expressed as the result of a differential operator $\O$ in 
$\ptl/\ptl K$
acting on $\exp(-\xi)$.
But then,
$$
  \int p(\xi)\exp(-\xi) \phi(x) dx = 
  \int \O \exp(-\xi) \phi(x) dx = 
  \O \int \exp(-\xi) \phi(x) dx
  \ideq \O J
  \ .
$$
We want to choose $\O$ so that this last expression yields the desired
results for test functions that are polynomials in
the coordinates $x^\mu$ of degree two
or less.
But the integral $J$ has a very simple form for such $\f$.  Up to
contributions that are negligible for large $K$,
it is just
a linear combination of terms of the form
$1/K^n$ or $\log{K}/K^n$.   Moreover the only monomials that yield
logarithmic terms are (in $\Mink^2$) $\f=t^2$, $\f=x^2$, and  $\f=1$.
In particular the monomials whose D'Alembertian vanishes produce only 
$1/K$, $1/K^2$ or $O(1/K^3)$, 
with the exception of $\f=1$, 
which produces 
a term in
$\log{K}/K$.  
These are the monomials that 
we 
don't want to
survive in $\O{J}$.  On the other hand $\f=t^2$ and $\f=x^2$ both
produce the logarithmic term $\log{K}/K^2$, and we do want them to
survive.
Notice further, that the survival of {\it any} logarithmic terms would
be bad, because, for dimensional reasons, they would necessarily bring
in an ``infrared'' dependence on the overall size of the region of
integration.  Taking all this into consideration, what we need from the
operator $\O$ is that it remove the logarithms and annihilate the
terms $1/K^n$.  Such an operator is
$$
   \O = \half (H+1)(H+2)   \qquad {\rm where} \quad H=K {\ptl\over\ptl K}
$$
is the homogeneity operator.
Applying this to $\exp(-\xi)$ turns out to yield precisely the
polynomial $p(\xi)$ that we were led to above in another manner,
explaining 
in a sense
why this particular polynomial arises.  
(The relation to
the binomial coefficients, traces back to an identity, proved by Joe
Henson, that expresses $(H+1)(H+2)...(H+n)\exp(-K)$ in terms of binomial
coefficients.)  Notice finally that $(H+1)(H+2)$ does {\it not}
annihilate $\log{K}/K$; but it converts it into $1/K$, 
which can be canceled by adding 
a delta-function to the integral kernel, as in fact we
did.  (It could also have been removed by a further factor of $(H+1)$.)

The situation for $\Mink^4$ is very similar to that for  $\Mink^2$.
The low degree monomials again produce terms in $1/K^n$ or
$\log{K}/K^n$,
but everything has an extra factor of $1/K$.  
Therefore
$\O={1\over6}(H+1)(H+2)(H+3)$
is a natural choice and leads to a polynomial based on the binomial
coefficients of $(1-1)^3$ instead of $(1-1)^2$.
From it we can derive both a causet D'Alembertian and a 
nonlocal, retarded deformation of the continuum D'Alembertian, as before.
It remains to be confirmed, however, that 
these expressions
 enjoy all the advantages
of the two-dimensional operators discussed above.
It also remains to be confirmed that these advantages persist in the
presence of curvature (but not, of course, curvature large compared to
the nonlocality scale $K$ that one is working with).

It seems likely that the same procedure would yield candidates
for retarded D'Alem\-bertians in all other spacetime dimensions.

\section{Continuous nonlocality, Fourier transforms and Stability}
In the course of the above reflexions, we have encountered some
D'Alembertian matrices for the causet and we have seen that
 the most promising among them
contain
a free parameter $K$ representing an effective nonlocality scale or
``meso-scale'', as I will sometimes call it.  For processes occurring on
this scale (assuming it is much larger than the ultraviolet scale $\UV$
so that a continuum approximation makes sense) one would expect to
recognize an effective nonlocal theory corresponding to the retarded
two-point function $\Bbar_K(x,y)$.
For clarity of notation, I will call the corresponding operator on
scalar fields ${\boxK}$, rather than $\Bbar_K$.  

Although 
its
nonlocality 
stems from the discreteness of
the underlying causet, 
${\boxK}$ 
is a perfectly well defined operator in the
continuum, which can be studied for its own sake.  At the same time, it
can help shed light on some questions that arise naturally in relation
to its causet cousin $B_{xy}$.  

One such question (put to me by Ted Jacobson) asks whether the evolution
defined in the causet by $B_{xy}$ is stable or not.  This seems
difficult to address
as such
 except by computer simulations, but if we transpose it to a
continuum question about  $\boxK$, we can come near to a full
answer. 
Normally, one expects that if there were an instability then $\boxK$
would possess an ``unstable mode'' (quasinormal mode), that is, a
spacetime function $\f$ of the form
$\f(x)=\exp(ik\cdot x)$ satisfying  $\boxK\f=0$, with the imaginary
part of the wave-vector $k$ being future-timelike.\footnote{$^\dagger$}
{One might question whether  $\boxK\f$ is defined at all for a general mode since
 the integral that enters into its definition might diverge, but for a
 putative unstable mode, this should not be a problem because the
 integral has its support precisely where the mode dies out: toward the
 past.}

Now by Lorentz invariance, $\boxK\f$ must be expressible in terms of 
$z=k\cdot k$, and it is not too difficult
to reduce it to an ``Exponential integral'' Ei in $z$. 
This being done,
some exploration in Maple strongly suggests
that the only solution of $\boxK\exp(ik\cdot x)=0$
is $z=0$, which would mean
the dispersion relation was unchanged from the usual one,
$\omega^2=k^2$.
If this is so, 
then 
no instabilities can result from 
the introduction of our nonlocality scale $K$, 
since the solutions of $\boxK\phi=0$ are
precisely those belonging to the usual D'Alembertian.
The distinction between 
propagation based on the latter and 
propagation based on $\boxK$ 
would repose only on 
the different relationship that $\f$ would bear to its sources;
propagation in empty (and flat) space would show no differences at all.
(The massive case might tell a different story, though.)

\subsection {Fourier transform methods more generally}

What we've just said is essentially that the Fourier transform of 
$\boxK$ vanishes nowhere in the complex $z$-plane ($z\ideq k\cdot k$), 
except at the origin.  
But this draws our attention to the Fourier transform
as yet another route for arriving at a nonlocal 
D'Alembertian.  
Indeed, most people investigating deformations of $\dal$ seem to have
thought of them in this way, including for example [6].
They
 have written down expressions like
$\dal\exp(\dal/K)$, but without seeming to pay much attention to whether
such an expression makes sense in a spacetime whose signature has not
been Wick rotated to $(++++)$.  
In contrast, the operator $\boxK$ of this paper was defined directly in
``position space''
as an integral kernel, not 
as
a formal function of $\dal$.
Moreover, because it is retarded,
its Fourier transform is rather special \dots.
By continuing in this vein, one can come up with a third derivation of
$\boxK$ as 
(apparently) 
the simplest operator whose Fourier
transform obeys the analyticity and boundedness conditions required in
order that $\boxK$ be well-defined and retarded.

The Fourier transform itself can be given in many forms, but the
following is among the simplest:
$$
  \boxK e^{ik\cdot x}|_{x=0} 
  \ = \ 
  {2z\over i}
  \int_0^\infty dt \; {e^{itz/K} \over (t-i)^2}
  \eqno(8)
$$
where
here,
$z=-k\cdot k/2$.

It would be interesting to learn what operator would result if one
imposed ``Feynman  boundary conditions'' on the inverse Fourier transform
of this function, instead of ``causal'' ones.

\section{What next?}
Equations (7) and (6) 
offer us
two distinct, but closely related, versions of $\dal$,
one suited to a causet and the other being an effective 
continuum operator arising as an average or limit of the first. 
Both are retarded and each is Lorentz invariant in the relevant sense.
How can we use them?
First of all, 
we can
take up the 
questions about wave-propagation raised
in the introduction, 
looking in particular for deviations from the
simplified model of [11]
based on ``direct transmission'' from source to sink
(a model that has much in common with the approach discussed above under
the heading ``Approach through the Green function'').
Equation (7) in particular, would let us propagate a
wave-packet through the causet and look for some of the effects
indicated in the introduction, like ``swerves'', scattering and
extinction. 
These of course 
hark back
directly to the granularity of the causet, but
even in 
the continuum 
limit
the nonlocality associated with 
(6) might modify the field emitted by a 
given 
source in
an interesting manner;
and this would be relatively easy to analyze. 

Also relatively easy to study would be the effect of the nonlocality on
free propagation in a curved background.  
Here one {\it would} expect some change to the
propagation law.
Because of the retarded character of $\boxK$, one
might also expect to see some sort of induced CPT violation in an expanding
cosmos.  Because (in a quantal context) this would disrupt the equality
between the masses of particles and antiparticles, it would
be a potential source of baryon-anti-baryon asymmetry not resting on any
departure from thermal equilibrium.

When discreteness combines with spacetime curvature, new issues arise. 
Thus, propagation of wave-packets in an expanding universe and in a
black hole background both raise puzzles having to do with the extreme
red shifts that occur in both situations
(so-called transplanckian puzzles).  
In the black hole context, the red shifts are of course responsible
for Hawking radiation, but their analysis in the continuum seems to
assign a role to 
modes of exponentially high frequency 
that arguably
should  be eschewed if one posits a minimum length.
Equation (7) offers a framework in which 
such questions
 can be
addressed without infringing on Lorentz invariance.
The same holds for questions about what happens to wave-packets in (say)
a de Sitter spacetime when they are traced backward toward the past far
enough so that their frequency (with respect to some cosmic rest frame)
exceeds Planckian values.  
Of course, such questions will not be resolved fully 
on the basis of classical equations of motion.
Rather one will have to formulate 
quantum field theory on a causet, 
or possibly one will have to go all the way to 
a quantal field on a quantal causet 
(i.e. to quantum gravity).
Nevertheless, a better understanding of the classical case is 
likely
to be relevant. 

I will not try to discuss here how to do quantum field theory on a
causet, 
or even in Minkowski spacetime with a nonlocal D'Alembertian. 
That would raise a whole set of new issues,
path-integral vs. operator methods and the roles of unitarity and causality
being just some of them.\footnote{$^\flat$}
{I will however echo a comment made earlier: I suspect that one should
 not try to formulate a path-integral propagator as such; rather one
 will work with Schwinger-Kel'dysh paths.  }
But it does seem in harmony with the aim of this paper to comment
briefly on the role of nonlocality in this connection.  
As we have seen, the ansatz (6) embodies a nonlocal
interaction that has survived in the continuum limit, and thus might be 
made the basis of a nonlocal field theory of the sort that people have
long been speculating about.

What is 
especially
interesting from this point of view is the potential for a new
approach to renormalization theory (say in flat spacetime $\Mink^d$).
People have sometimes hoped that nonlocality would eliminate the
divergences of quantum field theory, 
but as far as I can see, the opposite is true,
at least for the specific sort of nonlocality embodied in (6).
In saying this, I'm assuming that the divergences can all be traced to
divergences of the Green function $G(x-y)$ 
in the coincidence limit $x=y$.  
If this is correct then one would need to soften
the high frequency behavior of $G$,
in order to eliminate them.
But a glance at (8)
reveals that $\boxK$ has a milder ultraviolet behavior than $\dal$,
since its Fourier transform goes to a constant at $z=\infty$, rather
than blowing up linearly.  
Correspondingly, 
one would expect its Green
function to be
more singular
 than that of the local operator $\dal$,
making the divergences worse, not better.
If so, then 
one must look to the discreteness itself to cure the divergences; 
its associated nonlocality will not do the job.

But if nonlocality alone 
cannot remove the need for renormalization altogether,
it might 
nevertheless
open up a new and more symmetrical way to arrive at finite answers.
The point is that (8) behaves at $z=\infty$ like $1+O(1/z)$,
an expression whose reciprocal has exactly the same behavior!
The resulting Green function should therefore also be the sum of a
delta-function with a regular\footnote{$^\star$}
{At worst, it might diverge logarithmically on the light cone, but in
 that case, the residual divergence could be removed by adjusting the
 Fourier transform to behave like $1 + O(1/z^2)$.}
function
(and the same reasoning would apply in four dimensions).
The resulting 
Feynman diagrams would be finite 
{\it except for} contributions from the delta-functions.
But these could be removed by hand
(``renormalized away'').
If this idea worked out, it could provide a new approach to renormalization
based on a new type of Lorentz invariant regularization.  
(Notice that this all makes sense in real space, without the need for
Wick rotation.)

\section{How big is $\meso$?} 
From a phenomenological perspective, the most burning question is one
that I cannot really answer here: Assuming there are nonlocal effects
of the sort considered in the preceding lines, on what length-scales
would they be expected to show up?
In other words, what is the value of $\meso=K^{-1/2}$~?
Although I don't know how to answer this question theoretically,\footnote{$^\dagger$}
{The question of why $\UV$ and $\meso$ are not comparable joins the
 other ``large number'' (or ``hierarchy'') puzzles.  Perhaps their ratio
 is set dynamically (e.g. ``historically'' in relation to the large age
 of the cosmos), along with the small size of the cosmological constant
 and the large size of the cosmic diameter.  Such a mechanism could be
 either complementary to that suggested in [12] (explaining why the value
 about which Lambda fluctuates is so close to zero) or alternative to
 it.}
it is possible to deduce bounds on $\meso$ if we assume that the
fluctuations in individual values of 
$\dal\phi(causet)=B_K\phi$ are small, as
discussed above.
Whether such an assumption will still seem necessary at the end of the
day is of course very much an open question.  Not only could a sum over
individual elements of the causet counteract the fluctuations (as
already mentioned), but the same thing could result from the sum over
causets implicit in quantum gravity.  
This would be a sum of exponentially more terms, and as such it could
potentially remove the need for any intermediate nonlocality-scale
altogether. 

In any case, if we do demand that the fluctuations be elementwise
small, 
then
$\meso$ is bounded from below by this requirement.  (It is of course
bounded above by the fact that --- presumably --- we have not seen it
yet.)  Although this bound is not easy to analyze, a very crude estimate
that I will not reproduce here suggests that we make a small fractional
error in $\dal\phi$ when (in dimension four)
$$
               \lambda^2 \UV^2 \IR \ll \meso^5  \ ,
$$
where $\lambda$ is the characteristic length-scale associated with the
scalar field.
On the other hand, even the limiting continuum expression 
$\boxK\phi$ will be a bad approximation unless
$\lambda\gg\meso$.
Combining these inequalities yields 
$\lambda^2\UV^2\IR\ll\meso^5\ll\lambda^5$,
or $\UV^2\IR\ll\lambda^3$.
For smaller $\lambda$, accurate approximation to 
$\dal\phi$ is incompatible with small fluctuations.
Inserting for $\UV$ the Planck length\footnote{$^\flat$}
{This could be an underestimate if a significant amount of
 coarse-graining of the causet were required for spacetime to emerge.}
of $10^{-32}cm$ and for $\IR$ the Hubble radius, yields 
$\lambda\sim10^{-12}cm$
as the smallest wavelength that would be immune to 
the nonlocality.
That this is not an extremely small length,
poses the question 
whether observations already exist that could rule out nonlocality on
this scale.\footnote{$^\star$}
{Compare the interesting observations (concerning ``swerves'') in
 [13]}


\bigskip
\noindent
It's a pleasure to thank Fay Dowker and Joe Henson for extensive
discussions and help on these matters, during their visits to Perimeter
Institute. 
Research at Perimeter Institute for Theoretical Physics is supported in
part by the Government of Canada through NSERC and by the Province of
Ontario through MRI.
This research was partly supported
by NSF grant PHY-0404646.


\pagebreak
\phantom{blah blah}
\vskip 2 truein
\epsfxsize=6.2in
\FigureNumberCaption {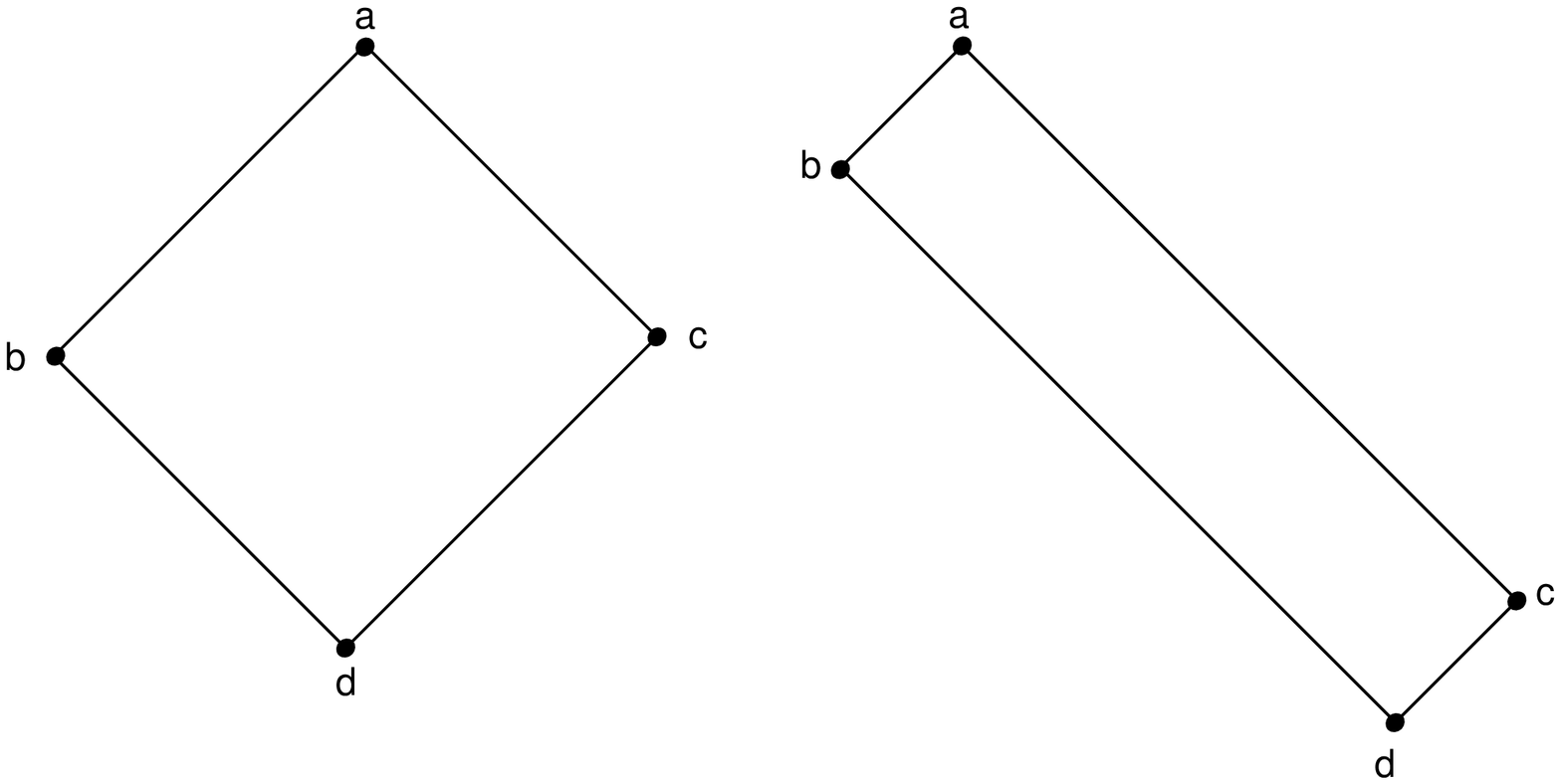} {1} {How a miracle might happen}

\vskip 2 truein

\epsfxsize=3.6in
\FigureNumberCaption{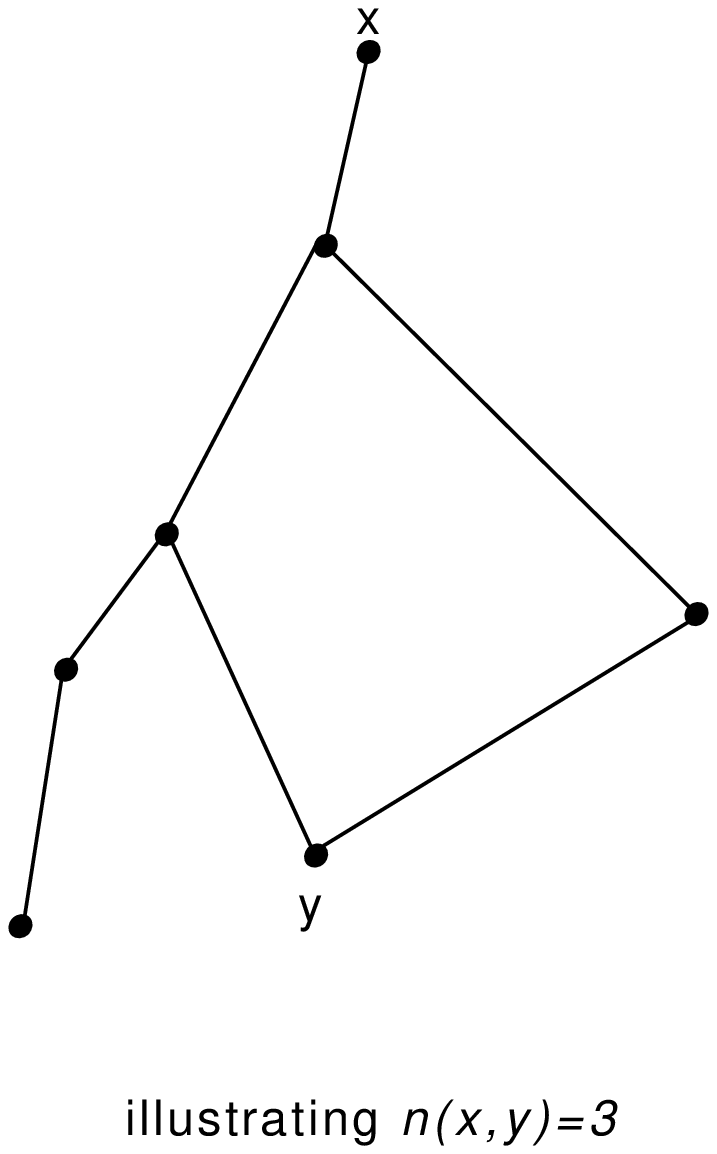}{2}%
{Illustration of the definition of the ``proximity measure'' $n(a,b)$.
 For the causet shown in the figure, $n(x,y)=3$ 
 because three elements intervene causally between $y$ and $x$.
 The first layer below element $x$ contains a single element, while the
 second, third and fourth, contain 2, 1, and 2, elements respectively.}

\pagebreak

\ReferencesBegin                             

\ref [1] 
Fay Dowker, Joe Henson and Rafael D.~Sorkin,
``Quantum Gravity Phenomenology, Lorentz Invariance and Discreteness'',
\journaldata {Modern Physics Letters~A} {19} {1829-1840} {2004}
\eprint{gr-qc/0311055}

\ref [2] 
R.M Dudley,
``Lorentz-invariant Markov processes in relativistic phase space'',
\journaldata{Arkiv f{\"o}r Matematik}{6(14)}{241-268}{1965}

\ref [3] 
Luca Bombelli, Joe Henson and Rafael D. Sorkin,
``Discreteness without symmetry breaking: a theorem''
(in preparation)


\ref [4]
See the article by Joe Henson in this volume (\eprint{gr-qc/0601121}).
Some other references of a general nature are: \linebreak
Luca Bombelli, Joohan Lee, David Meyer and Rafael D.~Sorkin, 
``Spacetime as a Causal Set'', 
  \journaldata {Phys. Rev. Lett.}{59}{521-524}{1987}; \lbr
Rafael D.~Sorkin,
``Causal Sets: Discrete Gravity (Notes for the Valdivia Summer School)'',
in {\it Lectures on Quantum Gravity}
(Series of the Centro De Estudios Cient{\'\i}ficos),
proceedings of the Valdivia Summer School, 
held January 2002 in Valdivia, Chile, 
edited by Andr{\'e}s Gomberoff and Don Marolf 
(Plenum, 2005)
\eprint{gr-qc/0309009}; and \linebreak
Fay Dowker, ``Causal sets and the deep structure of Spacetime'', 
in
{\it 100 Years of Relativity - Space-time Structure: Einstein and Beyond}"
ed Abhay Ashtekar 
(World Scientific, to appear)
\eprint{gr-qc/0508109}

\ref [5] 
John Collins, Alejandro Perez, Daniel Sudarsky, Luis Urrutia, and He'ctor Vucetich,
``Lorentz invariance and quantum gravity: an additional fine-tuning problem?''
gr-qc/0403053

\ref [6] 
K. Namsrai,
{\it Nonlocal Quantum Field Theory and Stochastic Quantum Mechanics}
(D. Reidel, 1986); 
J.W. Moffat, ``Finite nonlocal gauge field theory'',
\journaldata{Phys. Rev. D}{41}{1177-1184}{1990}

\ref [7] 
See the article by Fotini Markopoulou in this volume;
also
Sundance O. Bilson-Thompson, Fotini Markopoulou and Lee Smolin,
``Quantum gravity and the standard model'',
\eprint{hep-th/0603022}

\ref [8] 
Antony Valentini,
``Black Holes, Information Loss, and Hidden Variables''
\lbr
\eprint{hep-th/0407032}

\ref [9] 
Alan R. Daughton,			
 {\it The Recovery of Locality for Causal Sets and Related Topics},
  Ph.D. dissertation (Syracuse University, 1993)

\ref [10] 
Roberto  Salgado,			
``Toward a Quantum Dynamics for Causal Sets''
  (Ph.D. dissertation, Syracuse University, in preparation)

\ref [11] 
Fay Dowker, Joe Henson and Rafael D.~Sorkin,
``Wave propagation on a causet I: direct transmission along causal links''
(in preparation)

\ref [12]
Maqbool Ahmed, Scott Dodelson, Patrick Greene and Rafael D.~Sorkin,
``Everpresent $\Lambda$'',
\journaldata {Phys. Rev.~D} {69} {103523} {2004}
\eprint{astro-ph/0209274}

\ref [13]  
Nemanja Kaloper and David Mattingly,
``Low energy bounds on Poincar{\'e} violation in causal set theory'',
\eprint{astro-ph/0607485}

\end               


(prog1    'now-outlining
  (Outline 
     "\f......"
      "
      "
      "
   ;; "\\\\message"
   "\\\\Abstrac"
   "\\\\section"
   "\\\\subsectio"
   "\\\\appendi"
   "\\\\Referen"
   "\\\\ref....[^|]"
  ;"\\\\ref....."
   "\\\\end

%% file: mathmacros.tex





\font\openface=msbm10 at10pt
 %

\def\Minkowski     {{\hbox{\openface M}}}
\def\Euclid        {{\hbox{\openface E}}}

 %
 %
 %



\font\german=eufm10 at 10pt

\def\Buchstabe#1{{\hbox{\german #1}}}










%

%
%



%



\def\sqr#1#2{\vcenter{
  \hrule height.#2pt 
  \hbox{\vrule width.#2pt height#1pt 
        \kern#1pt 
        \vrule width.#2pt}
  \hrule height.#2pt}}


\def\dal{\mathop{\,\sqr{7}{5}\,}}
\def\block{\dal}



\def\lto{\mathop
        {\hbox{${\lower3.8pt\hbox{$<$}}\atop{\raise0.2pt\hbox{$\sim$}}$}}}
\def\gto{\mathop
        {\hbox{${\lower3.8pt\hbox{$>$}}\atop{\raise0.2pt\hbox{$\sim$}}$}}}
%
%
%


\def\half{{1 \over 2}}


\def\part{\subseteq}		

\def\ptl{\partial}

\def\braces#1{ \{ #1 \} }



\def\to{\mathop\rightarrow}	

\def\ideq{\equiv}		



\def\interior #1 {  \buildrel\circ\over  #1}     



\def\grad{\nabla}


\def\basisvector#1#2#3{
 \lower6pt\hbox{
  ${\buildrel{\displaystyle #1}\over{\scriptscriptstyle(#2)}}$}^#3}

\def\eps{\varepsilon}


\def\bar{\overline}		






%% file: mathmacros.greekbold.tex

%






\font\bmit=cmmib10			
\font\expo=cmmib10 at 10 true pt	

\newfam\boldmath

\textfont8=\bmit 
\scriptfont8=\expo 
\scriptscriptfont8=\expo

  \mathchardef\alpha="710B     
  \mathchardef\beta="710C
  \mathchardef\gamma="710D     
  \mathchardef\delta="710E
  \mathchardef\epsilon="710F   
  \mathchardef\zeta="7110
  \mathchardef\eta="7111       
  \mathchardef\theta="7112
  \mathchardef\iota="7113
  \mathchardef\kappa="7114     
  \mathchardef\lambda="7115
  \mathchardef\mu="7116        
  \mathchardef\nu="7117
  \mathchardef\xi="7118        
  \mathchardef\pi="7119
  \mathchardef\rho="711A       
  \mathchardef\sigma="711B
  \mathchardef\tau="711C       
  \mathchardef\upsilon="711D
  \mathchardef\phi="711E
  \mathchardef\chi="711F
  \mathchardef\psi="7120
  \mathchardef\omega="7121     
  \mathchardef\varepsilon="7122
  \mathchardef\vartheta="7123
  \mathchardef\varpi="7124
  \mathchardef\varrho="7125
  \mathchardef\varsigma="7126
  \mathchardef\varphi="7127

  \mathchardef\imath="717B	
  \mathchardef\jmath="717C	
  \mathchardef\ell="7160
  \mathchardef\partial"7140



\def\goesto#1{\quad\lower 1ex\overrightarrow{\ssize\hphantom{M}
    #1 \hphantom{M}}\quad}

\def\goesblank{{\;\hbox to 25pt{\rightarrowfill}\;}}


%% file: msmacros.tex

%



%
%
%
%
%
%
%
%
%
%
%
%

%
 \let\miguu=\footnote
 \def\footnote#1#2{{$\,$\parindent=9pt\baselineskip=13pt%
 \miguu{#1}{#2\vskip -7truept}}}
%
%

\def\linebreak{\hfil\break}
\def\lbr{\linebreak}
\def\pagebreak{\vfil\break}


\def\BulletItem #1 {\item{$\bullet$}{#1 }}
\def\bulletitem #1 {\BulletItem{#1}}

\def\AbstractBegins
{
 \singlespace                                        
 \bigskip\leftskip=1.5truecm\rightskip=1.5truecm     
 \centerline{\bf Abstract}
 \smallskip
 \noindent	
 } 
\def\AbstractEnds
{
 \bigskip\leftskip=0truecm\rightskip=0truecm       
 }

\def\ReferencesBegin
{
 \singlespace					   
 \vskip 0.5truein
 \centerline           {\bf References}
 \par\nobreak
 \medskip
 \noindent
 \parindent=2pt
 \parskip=6pt			
 }
 %


\def\section #1 {\bigskip\noindent{\headingfont #1 }\par\nobreak\noindent}

\def\subsection #1 {\medskip\noindent{\subheadfont #1 }\par\nobreak\noindent}
%

\def\reference{\hangindent=1pc\hangafter=1} 

\def\ref{\reference}

 %

\def\journaldata#1#2#3#4{{\it #1\/}\phantom{--}{\bf #2$\,$:} $\!$#3 (#4)}
 %

\def\eprint#1{{\tt #1}}
 %
 %
 %


\def\author#1 {\medskip\centerline{\it #1}\bigskip}

\def\address#1{\centerline{\it #1}\smallskip}

\def\furtheraddress#1{\centerline{\it and}\smallskip\centerline{\it #1}\smallskip}

\def\email#1{\smallskip\centerline{\it address for email: #1}}

\def\PrintVersionNumber{
 \vskip -1 true in \medskip 
 \rightline{version \Version} 
 \vskip 0.3 true in \bigskip \bigskip}



\font\titlefont=cmb10 scaled\magstep2 

\font\headingfont=cmb10 at 12pt
%

\font\subheadfont=cmssi10 scaled\magstep1 
%





